\documentclass[authoryear,preprint,review,12pt]{elsarticle}

\usepackage{amsmath}
\usepackage{amssymb}
\usepackage{graphicx}
\usepackage{color}
\usepackage{subfigure}

\usepackage{url}

\begin{document}

\begin{frontmatter}

\title{Btrim: A fast, lightweight adapter and quality trimming program 
for next-generation sequencing technologies}
\author{Yong Kong}
\address{
Department of Molecular
Biophysics and Biochemistry\\
W.M. Keck Foundation Biotechnology Resource Laboratory \\
Yale University\\
333 Cedar Street, New Haven, CT 06510, USA \\
Phone: 1-203-737-5011 Fax: 1-203-785-6404\\
\ead{yong.kong@yale.edu}
}

\begin{abstract}
Btrim is a fast and lightweight software to trim adapters and low quality
regions in reads from ultra high-throughput next-generation sequencing
machines.
It also can reliably identify barcodes and
assign the reads to the original samples. 
Based on a modified Myers's bit-vector dynamic programming algorithm,
Btrim can handle indels in adapters and barcodes.
It removes low quality regions and trims
off adapters at both or either end of the reads. 
A typical trimming of 30M reads with two sets of adapter pairs
can be done in about a minute with a small memory footprint.
Btrim is a versatile stand-alone tool that
can be used as the first step in virtually all next-generation 
sequence analysis pipelines. 
The program is available at \url{http://graphics.med.yale.edu/trim/}.

\end{abstract}

%\bigskip


\end{frontmatter}

\newpage

\section{Introduction}

Next-generation sequencing technologies generate huge amount of sequences
in an ever increasing speed.
Some of these reads contain adapters and other exogenous contents 
by experimental designs.
On other situations, 
adapters were sequenced inadvertently when they should not 
due to operational errors and other unknown reasons.
If these adapters were not trimmed out, they would interfere with the
downstream data analysis, such as mapping the reads to the reference genome
and \emph{de novo} assembly.
Unlike the traditional homology search tools such as BLAST,
nearly all the current popular mapping programs for short reads
from the next-generation sequencing technologies
use global matching algorithms,
trying to match the whole length of the read to the reference genome 
if the number of mismatches or
edit distance (mismatches and indels) is smaller than a pre-defined threshold. 
Since the adapters are not part of the biological sequences,
they are treated as matching errors and the reads containing them 
are not mapped.  
If the reads containing the exogenous contents are below the
error threshold and eventually mapped,
they will be the cause of many potential issues 
for downstream analysis.
For example,
the exogenous contents will increase the false positive rate 
for the subsequent variant analysis.
They will also cause problems for \emph{de novo} assembly.

For most of the next-generation sequencing technologies, 
the quality of the sequencing reads becomes lower towards the end of the
reads. If too many sequencing errors occur in the end of the reads,
it would cause problem for mapping and other downstream analysis,
for the same reason as mentioned above, 
even if the reads contain
high-quality and mappable bases in the beginning.  
As the length of sequences becomes longer,
this issue will become more serious.

Some experiments using the next-generation sequencing platforms
utilize barcode technology: distinct barcode sequence tag representing 
individual samples are ligated into the DNA sequences and the pooled samples
are sequenced. 
To assign the reads back to the original samples 
we need to reliably identify the barcodes within the sequence reads, 
in the presence of
possible experimental and sequencing errors in the barcodes.
In practice, barcode identification can be achieved by using
algorithms similar to those used in finding the adapters.

The similar problem of adapter and low-quality region trimming existed
for the traditional Sanger sequencing technology.  
Most people dealt with
the problem using some makeshift solutions without paying much
attention to their efficiency, 
which worked fine for small number of sequences.
With the advent of next-generation sequencing technologies, 
the huge volume of sequences generated asks for more efficient
software to handle the problem.
Some existing specialized packages, 
such as mapping program novoalign (\url{http://www.novocraft.com/}),
have some built-in adapter-removal functions.
A versatile stand-alone
program will be very useful for flexible downstream analysis
%(such as mapping to reference genome and de novo assembly)  
using different software packages.
For example,
after exogenous contents and low quality regions removed,
the processed sequences can then be used as input for 
different mapping or assembly programs to compare their performance.
On the other hand, 
a range of different parameters can be tried to trim the original sequences
and the optimal parameter can be chosen based on the output of downstream
analysis.

Here we present a fast and lightweight program
for adapter and quality trimming.
For adapter trimming,
the program is based on 
modified Myers's bit-vector dynamic programming algorithm~\cite{Myers1999}.  
For quality trimming, a simple moving window algorithm is used
and the reads are trimmed at the point where
the average quality score within the window drops below a threshold.

\section{Bit-vector dynamic programming algorithm for adapter trimming
and barcode identification}

As one of the fastest dynamic programming algorithms 
available with edit distance as
the error model (each mismatch, insertion, or deletion counts as one error),
the Myers's bit-vector dynamic programming algorithm finds all locations at which the query matches a
substring of the target sequence of length $n$ with $k$ or fewer errors.
The algorithm scales linearly with the length of the target sequence ($n$)
 when the
length of the query is less than the machine word size $w$
(typically, $w=32$ for 32-bit machines and $w=64$ for 64-bit machines),
regardless of $k$ or query length~\cite{Myers1999}.  

Before the search starts, the algorithm pre-processes the query sequences.
Since in most cases 
the number of query sequences is far smaller than that of the target sequences,
this pre-processing time is negligible.
After the query sequences have been set up, in the subsequent search phase
the algorithm scans the target sequences linearly and one-by-one, reporting
any occurrences of the query with $k$ or fewer errors.
Hence our trimming program has a time complexity $O(q n)$,
for $q$ queries (adapters or barcodes).

The original Myers's algorithm searches from the beginning of the
target sequence towards the end and only reports the \emph{stopping} position
of query occurrences in the target sequence.
For example,
if the query is ACGT and the target sequence is AT\textbf{AGT}CCGGA,
the position of the second ``T'' in the target sequence
is reported for an occurrence 
with one deletion.
In order to determine precisely the boundary between the insert and
the adapters or barcodes, the \emph{starting} position 
of query occurrences in the target sequence is needed 
(for example, when 3'-adapter is being trimmed).
In the previous example we need to know the position
of the second ``A''.
By algorithm's design no traditional trace-back could be done to
find the query's starting position since no standard dynamic programming
matrix is saved.
Furthermore a trace-back would defeat the efficiency of the algorithm.
To find query's starting positions, 
we modified Myers algorithm for the 3'-adapter search.
The modified algorithm
carries out the search backward along the target sequence.
In order to do so, the query sequences have to be set up backward also.

The parameter $k$, the upper limit of the number of match errors,
is set equal to $3$ as default for 5'-adapter and equal to $4$ for 3'-adapter,
but they can be changed to be based on the length of the adapters,
or as user-specified. 
In the default mode for adapter trimming,
for each pair of 5'- and 3'-adapters, the program searches for the locations
of 5'-adapter within the beginning region of the read (default 
$1.3 \times $ query length). 
If none of the 5'-adapters is found, the trimming is failed for the read.
If one or more of the 5'-adapters are found with $k$ or fewer errors,
the adapter with the smallest error is selected.
The corresponding 3'-adapter is then used as query to search 
the rest of the read, using the above-mentioned modified algorithm.

If the selected 3'-adapter is found in the region,
then the location with the smallest error is used to trim the read;
otherwise, quality scores of the read are used to trim the 3'-end.
To use the quality scores, the average score within a moving window is 
calculated. The window moves from the 5' trimming point to the end of
the read.  If the average score drops below a preset threshold,
then the 3' trimming point is found at the beginning of the moving window.

Other modes of adapter trimming are also available, such as
passing the read to quality trimming if the adapters cannot be found,
or trimming based on the 3'-adapters only.

One of the advantages of Myers's algorithm is that it can handle
ambiguous and wildcard letters effortlessly.  
This makes it more suitable for adapter trimming, 
since usually the adapters are designed with degenerate bases 
incorporated in specific positions.
To handle ambiguous bases, simply put them into square brackets,
such as [GC] for either G or C in the same position.
For wildcard letters (which match any bases in the position), use ``.''
in the query.
IUPAC codes can be incorporated easily.
By default the query search is case insensitive; it can be switched to
case sensitive by a command line option.

In addition to adapter trimming, Btrim can do the trimming based on
quality scores only.
Regardless of the trimming mode,
as an option,
all flanking fragments and their quality scores that have been trimmed off 
can be written into a separate file
with detailed trimming information for each read,
such as the trimming status, the adapters/barcodes found,
the errors for the adapter/barcode, and the location of them in the read.
This file can be parsed easily and may be useful
for downstream analysis, such as barcode assignment.
The details about how to use the program can be found in the document
in the program's web site.

\begin{figure}[!tpb]
  \centering
    \includegraphics[angle=270,width=\columnwidth]{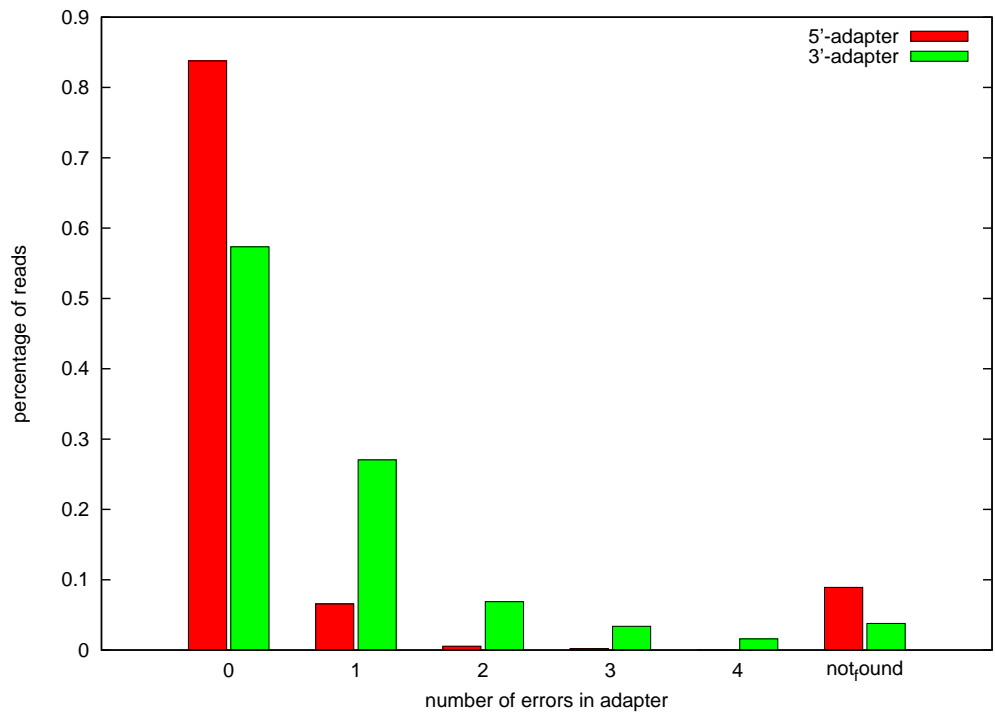}
  \caption{
    Error distribution of adapter trimming
for a typical 75bp Illumina lane.
The two pair of adapters used in trimming
are  agggaggacgatgcgg / gtgtcagtcacttccagcgg
and ccgctggaagtgactgacac / ccgcatcgtcctccct.
Default parameters are used.
  \label{F:err}
  }
\end{figure}

Figure~\ref{F:err} shows the error distribution of adapter trimming
for sequences from a typical Illumina lane. 
With two sets of 5'- and 3'-adapter pairs (each for one orientation of 
the insert), 
30M 75bp Illumina reads are processed
in 57 seconds in a machine with a 3.16GHz Intel Xeon processor.
As the target sequences are read in sequentially,
the memory footprint of the program is very small (less than 1M). 
Btrim supports FASTQ format currently (both Sanger and Illumina formats).
The program is written in C programming language and
available at \url{http://graphics.med.yale.edu/trim/}.

\section*{Acknowledgment}
  This work was supported by
  Yale School of Medicine.

\bibliographystyle{natbib}

%\bibliography{trim.bib}

%\begin{thebibliography}{1}
%
%\bibitem{Myers1999} 
%Myers, G (1999) 
%A fast bit-vector algorithm for approximate string matching based on 
%dynamic programming, {\em J. ACM}, {\bf 46}, 395-415.
%
%\end{thebibliography}

\end{document}